\documentclass[aps,pra,twocolumn,showpacs,amsmath,amssymb,preprintnumbers,superscriptaddress,10pt]{revtex4-1}

\usepackage{amsmath,amssymb}
\usepackage{bm}
\usepackage[latin1]{inputenc}
\usepackage{graphicx}
\usepackage{epstopdf}
\usepackage{SIunits}
\usepackage{hyphenat}

\usepackage{color}
\usepackage{pgfplots}
\usepgfplotslibrary{groupplots}
\usepackage{tabularx}
\usepackage{multirow}
\usepackage{natbib}

\usepackage{url}

\usepackage[bookmarks=false]{hyperref}
\usepackage[all]{hypcap}
\usepackage[capitalise]{cleveref}

\definecolor{pink}{RGB}{255,0,255}

\definecolor{dark-greenish-turquoise}{RGB}{0,160,110}

\definecolor{ruby}{RGB}{192,47,29}

\begin{document}

\title{Laser annealing heals radiation damage in avalanche photodiodes}

\author{Jin Gyu Lim}
\email{j29lim@uwaterloo.ca}
\affiliation{Institute for Quantum Computing, University of Waterloo, Waterloo, ON, N2L~3G1 Canada}
\affiliation{\mbox{Department of Electrical and Computer Engineering, University of Waterloo, Waterloo, ON, N2L~3G1 Canada}}

\author{Elena Anisimova}
\affiliation{Institute for Quantum Computing, University of Waterloo, Waterloo, ON, N2L~3G1 Canada}
\affiliation{Department of Physics and Astronomy, University of Waterloo, Waterloo, ON, N2L~3G1 Canada}

\author{Brendon L. Higgins}
\affiliation{Institute for Quantum Computing, University of Waterloo, Waterloo, ON, N2L~3G1 Canada}
\affiliation{Department of Physics and Astronomy, University of Waterloo, Waterloo, ON, N2L~3G1 Canada}

\author{Jean-Philippe Bourgoin}
\affiliation{Institute for Quantum Computing, University of Waterloo, Waterloo, ON, N2L~3G1 Canada}
\affiliation{Department of Physics and Astronomy, University of Waterloo, Waterloo, ON, N2L~3G1 Canada}

\author{Thomas Jennewein}
\affiliation{Institute for Quantum Computing, University of Waterloo, Waterloo, ON, N2L~3G1 Canada}
\affiliation{Department of Physics and Astronomy, University of Waterloo, Waterloo, ON, N2L~3G1 Canada}
\affiliation{Quantum Information Science Program, Canadian Institute for Advanced Research, Toronto, ON, M5G~1Z8 Canada}

\author{Vadim Makarov}
\affiliation{Department of Physics and Astronomy, University of Waterloo, Waterloo, ON, N2L~3G1 Canada}
\affiliation{Institute for Quantum Computing, University of Waterloo, Waterloo, ON, N2L~3G1 Canada}
\affiliation{\mbox{Department of Electrical and Computer Engineering, University of Waterloo, Waterloo, ON, N2L~3G1 Canada}}

\date{January 30th, 2017}

\begin{abstract}
Avalanche photodiodes (APDs) are a practical option for space-based quantum communications requiring single-photon detection. However, radiation damage to APDs significantly increases their dark count rates and reduces their useful lifetimes in orbit. We show that high-power laser annealing of irradiated APDs of three different models (Excelitas C30902SH, Excelitas SLiK, and Laser Components SAP500S2) heals the radiation damage and substantially restores low dark count rates. Of nine samples, the maximum dark count rate reduction factor varies between 5.3 and 758 when operating at $-80~\celsius$. The illumination power to reach these reduction factors ranges from $0.8$ to $1.6~\watt$. Other photon detection characteristics, such as photon detection efficiency, timing jitter, and afterpulsing probability, remain mostly unaffected. These results herald a promising method to extend the lifetime of a quantum satellite equipped with APDs.
\end{abstract}

\maketitle

\section{Introduction}
\label{sec:introduction}

Quantum communications protocols, such as quantum key distribution (QKD) \cite{bennett1984, ekert1991}, quantum teleportation \cite{bennett1993}, and Bell's inequality tests \cite{bell1964}, are limited to transmission distances of only a few hundred kilometers \cite{yin2016, ma2012.Nature-489-269, scheidl2010} under the restrictions of present technology. For global-scale quantum communications, quantum repeaters \cite{briegel1998} and quantum satellites \cite{buttler1998, rarity2002, aspelmeyer2003} are potential solutions. Unfortunately, quantum repeaters are not ready for deployment as quantum memories with sufficient storage times and fidelities, upon which quantum repeaters depend, are still being developed \cite{simon2010,sangouard2011}. On the other hand, quantum communications to satellite platforms are feasible today \cite{gilbert2000, bonato2009, ursin2009, meyer-scott2011, xin2011, nordholt2002, takenaka2011, higgins2012, vallone2015}, with China being the first country to successfully launch a quantum satellite \cite{gibney2016}.

One of many challenges in achieving long-distance quantum communications is the noise floor imposed by detector dark counts \cite{scarani2009}---false photon detection events caused by thermally excited, tunnelling, and trapped electrons \cite{haitz1965}. A previous study \cite{bourgoin2013.NewJPhys-15-023006} examined the performance of both downlink and uplink satellite quantum communication designs under various conditions. Uplink communication, where the detectors are placed on the satellite, is particularly interesting because of potentially simpler satellite designs and easy interchangeability of sources at the ground station---for this approach, QKD, quantum teleportation, and Bell tests perform well with a dark count rate up to about $200~\hertz$ per detector.

Silicon avalanche photodiodes (APD) are an appropriate choice for the single-photon detector on a satellite because of their low dark count rate, good sensitivity in $400$--$1000~\nano\meter$ wavelength range (covering wavelengths near $785~\nano\meter$ for optimal uplink transmissions \cite{bourgoin2013.NewJPhys-15-023006}), and no need for cryogenic cooling \cite{cova2004, hadfield2009, eisaman2011}. However, proton radiation in low Earth orbit significantly increases APD dark count rates over time \cite{sun1997,sun2001,sun2004, tan2013, tang2016}. In a recent experiment \cite{anisimova2015}, APD samples (Excelitas C30902SH, Excelitas SLiK, and Laser Components SAP500S2) were irradiated by different fluences of $106~\mega\electronvolt$ protons to simulate radiation effects over 0.6, 6, 12, and 24 months in a representative low Earth polar orbit. There it was shown that thermal annealing of irradiated APDs at up to $100~\celsius$ can repair some of the damage, resulting in up to 6.6-fold dark count rate reduction. A separate study showed that laser annealing can lower non-irradiated APD dark count rates by up to 5.4 times \cite{bugge2014}. 

Here we perform laser annealing on nine irradiated APDs. We find that laser annealing successfully decreases the dark count rates in all nine irradiated APD samples by a factor ranging from 5.3 to 758 when operated at $-80~\celsius$. We demonstrate dark count rate reductions due to laser annealing can exceed those from thermal annealing. Notably, we observe that dark count rates are reduced even when laser annealing is applied to APDs that were already thermally annealed. Laser annealing also affects other important photon counting parameters including photon detection efficiency, timing jitter, and afterpulsing probability, but the operation of quantum communications applications should not be significantly influenced by these changes. 

\section{Experimental setup}
\label{sec:setup}

\begin{table*}
	\caption{\label{tab:APDs-conditions-and-results}{\bf Summary of detector samples, applied radiation, previous thermal annealing, and measured results of laser annealing.} The detectors are referred to by the given sample IDs throughout the paper. The highest reduction factor is the ratio between the initial dark count rate and the lowest dark count rate after laser annealing---the corresponding laser power for this is also given.}
	\centering
	\renewcommand{\arraystretch}{1.45}
   \begin{tabular}{
   		>{\centering\arraybackslash}m{18mm}
    	 	>{\centering\arraybackslash}m{24mm}
    		>{\centering\arraybackslash}m{29mm}
    	 	>{\centering\arraybackslash}m{25mm}|
    	 	>{\centering\arraybackslash}m{14mm}
    		>{\centering\arraybackslash}m{12mm}
    		>{\centering\arraybackslash}m{14mm}
    	 	>{\centering\arraybackslash}m{16mm}
    	 	>{\centering\arraybackslash}m{13mm}}
    	\hline\hline
    \centering
	\multirow{3}{16mm}{\centering\textbf{Sample ID}} & \multirow{3}{25mm}{\centering\textbf{$\bm{106~\mega\electronvolt}$ proton fluence ($\bm{\centi\meter^{-2}}$)}} & \multirow{3}{22mm}{\centering\textbf{Equivalent time in 600~km polar orbit (months)}} & \multirow{3}{23mm}{\centering\textbf{Thermal annealing procedure}} & \multicolumn{3}{c}{\textbf{Dark count rate at $\bm{-80~\celsius$}}} & \multirow{3}{15mm}{\centering\textbf{Annealing power ($\bm{\watt}$)}} & \multirow{3}{13mm}{\centering\textbf{$\bm{V_\text{excess}}$ ($\bm{\volt}$)}}\\

		\cline{5-7}
	  & & & & Before (\hertz) & Lowest after (\hertz) & Highest reduction factor & \\

	\hline
	C30902SH-1 & $10^9$ & 6 & None & $347$ & $2.3$ & 150 & $0.8$ & 14\\
	C30902SH-2 & $10^9$ & 6 & None & $363$ & $2.64$ & 137 & $1.5$ & 14\\
	SLiK-1 & $10^8$ & 0.6 & $2~\hour~@~{+}100~\celsius$ & $6.71$ & $0.16$ & 41.7 & $1.4$ & 14\\
	SLiK-2 & $10^8$ & 0.6 & $2~\hour~@~{+}100~\celsius$ & $2.19$ & $0.42$ & 5.3 & $0.8$ & 14\\
	SLiK-3 & $4\times10^9$ & 24 & $4~\hour~@~{+}80~\celsius$, $2~\hour~@~{+}100~\celsius$ & $43.1$ & $2.09$ & 21 & $1.4$ & 14\\
	SLiK-4 & $10^9$ & 6 & None & $192$ & $8.3$ & 23 & $1.0$ & 20\\ 
	SLiK-5 & $4\times10^9$ & 24 (with bias voltage applied) & $3~\hour~@~{+}80~\celsius$, $2~\hour~@~{+}100~\celsius$ & $447$ & $58$ & 7.7 & $1.0$ & 20\\
	SAP500S2-1 & $4\times10^9$ & 24 & $4~\hour~@~{+}80~\celsius$, $2~\hour~@~{+}100~\celsius$ & $1579$ & $2.08$ & 758 & $1.4$ & 20\\
	SAP500S2-2 & $10^8$ & 0.6 & $2~\hour~@~{+}100~\celsius$ & $213$ & $1.66$ & 128 & $1.6$ & 20\\
\hline\hline

  \end{tabular}
\end{table*}

\subsection{Test samples}

We test the same Excelitas C30902SH, Excelitas SLiK, and Laser Components SAP500S2 devices used in the previous study \cite{anisimova2015}. These were chosen as they are the only commercially available thick-junction APD models (thick-junction APDs have higher detection efficiencies at the quantum signal wavelength around $785~\nano\meter$, which makes them the most appealing for use on a quantum satellite receiver). All our APD samples are hermetically sealed in glass-window packages and the photosensitive areas of C30902SH and SAP500S2 are $500~\micro\meter$ in diameter, while that of SLiK is $180~\micro\meter$ in diameter. \Cref{tab:APDs-conditions-and-results} presents the details of previous radiation testing and thermal annealing procedures performed on our test samples. These samples were stored in a $-20~\celsius$ freezer before our experiments to prevent any spontaneous annealing at room temperature. 

For every detector, the dark count rate after irradiation (and any thermal annealing) is so high that the devices are saturated when operating at room temperature. For this reason, all samples are characterized at $-80~\celsius$ in our cold-temperature characterization apparatus. However, unlike the other devices, each Excelitas SLiK comes with a built-in thermistor and thermoelectric cooler (TEC), allowing additional testing to be performed in-situ in our laser annealing apparatus at an operating temperature of $-30~\celsius$.

\begin{figure}[h]
\centering
\includegraphics[width=0.73\columnwidth]{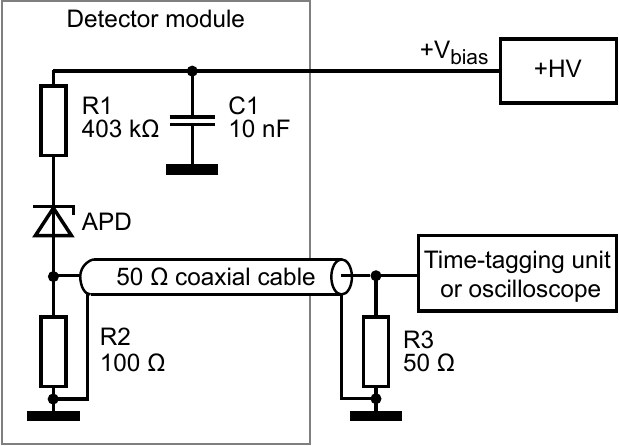}
\caption{\label{fig:detector-module}{\bf Detector module circuit.} Each detector module houses a number of APDs (up to six), all of which were irradiated to a common fluence. The circuit diagram for a single detector channel is shown---each APD is embedded in passive avalanche quenching electronics with an external high-voltage supply and output to a time-tagging unit (the time-tagging unit has a fixed-threshold discriminator at each input) or to an oscilloscope.}
\end{figure}

\begin{figure*}
\centering
\includegraphics[width=\textwidth]{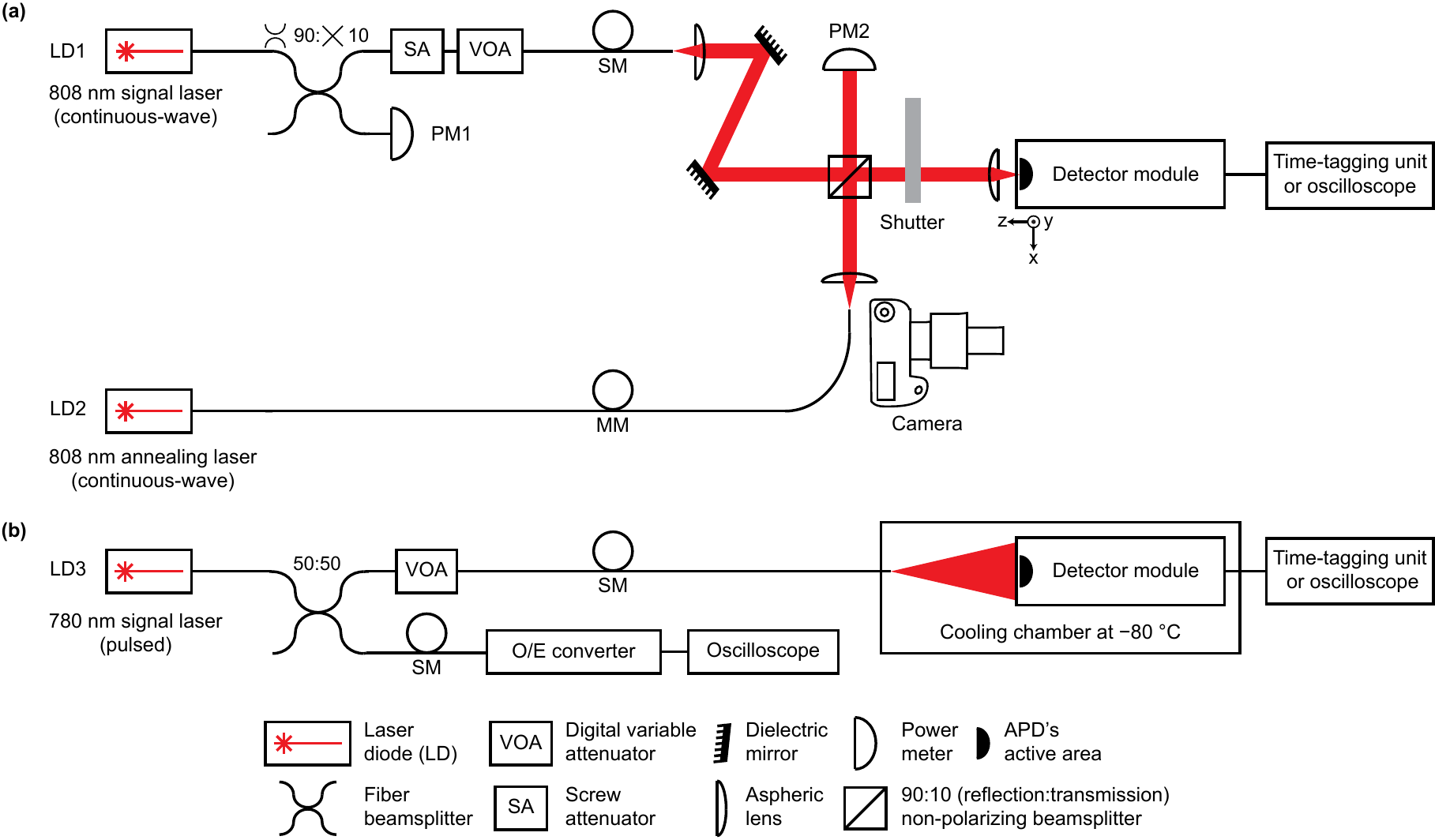}
\caption{\label{fig:experimental-setup}{\bf Experimental setup.} (a) Setup for laser annealing and characterization at room temperature. (b) Setup for characterization at $-80~\celsius$. In (a), LD1 and LD2 are overlapped such that high-power laser from LD2 can be focused on the APD's active area by focusing LD1 onto it. LD1 is also used for measuring SLiKs' photon detection efficiency. The detector module is mounted on an XYZ translation stage, which can move between the laser setup and camera. The free-space portion of the setup is enclosed in a light-tight box. In (b), diverging weak coherent pulses are applied on APDs for full characterization. SM: single-mode optic fiber; MM: multi-mode optic fiber; O/E: optical-to-electrical. }
\end{figure*}

The samples are held in detector modules, which are moved between the annealing and characterization apparatuses as necessary and provide avalanche detection and quenching.
We employ a passive quenching circuit \cite{haitz1965, cova1996} (see \cref{fig:detector-module}) for processing output avalanche pulses. In this circuit, the APD is reverse-biased in Geiger-mode \cite{haitz1965}, where the bias voltage ($V_\text{bias}$) is set above the detector's breakdown voltage ($V_\text{br}$) and the APD becomes sensitive to single photons \cite{cova1996}. $V_\text{excess}~(= V_\text{bias} - V_\text{br})$, typically ${\sim}20~\volt$, generates a high electric field in the detector's depletion and avalanche regions. When a detection takes place (ideally, by a photon incident on the APD active area), an electron-hole pair is generated. The high electric field in the p-n junction causes impact ionization and produces an avalanche current flow. The current flow is detected as a voltage drop across R2 in parallel with a $50~\ohm$ impedance coaxial cable. If this voltage drop is greater than a discriminator's fixed threshold voltage, a detection event is recorded. R1 quenches the current flow by lowering the voltage across the APD close to $V_\text{br}$ \cite{cova1996}. Once the diode voltage reaches near $V_\text{br}$ and the steady-state current flow ($V_\text{excess}/R1$) is below a latch current of ${\sim}100~\micro\ampere$, the avalanche current flow stops \cite{cova1996}. Our detector module has the steady-state current of ${\approx}50~\micro\ampere$ for $V_\text{excess} = 20~\volt$. 

We measure the dark count rate, relative changes in photon detection efficiency ($P_\text{de}$) (and absolute $P_\text{de}$ for SLiKs characterised at $-30~\celsius$), and afterpulsing time distribution through the analysis of detection event times produced by a time-tagging unit with $156.25~\pico\second$ resolution (UQDevices 16-channel model). Timing jitter ($\Delta t_\text{jitter}$) is measured using an oscilloscope (LeCroy 640Zi).

\subsection{Laser annealing apparatus}

\Cref{fig:experimental-setup}(a) shows the laser annealing apparatus, which improves upon the laser annealing experimental setup of Ref. \onlinecite{bugge2014}. Our apparatus allows us to laser-anneal APDs, take pictures of their active areas, measure their dark count rates and $P_\text{de}$, and scan $P_\text{de}$ across the entire active area to check whether laser annealing has produced any local damage. 

Our setup consists of a single-mode (SM) continuous-wave $808~\nano\meter$ signal laser (QPhotonics QFLD-808-100S), and one multi-mode (MM) $0$--$30~\watt$ continuous-wave annealing laser LD2 (Jenoptik JOLD-30-FC-12). The pigtail of the signal laser LD1 is connected to a 90:10 fiber beamsplitter (Thorlabs FC780-90B FC). One output port is connected to a power meter (PM1; OZ Optics POM-300-VIS), while the other output port is connected to two attenuators in series: a screw attenuator reduces laser power to the $\nano\watt$ range, and a digital variable attenuator (OZ Optics DA-100-35-770/830/850-5/125-5-40-LL) then brings laser illumination down to the single-photon level. The degree of attenuation and the output power from the weak coherent continuous-wave light source are calibrated to apply the mean photon count rate of $48.8~\kilo\hertz$ at the sample. The attenuated continuous-wave laser beam is sent to a collimation setup in SM fiber, then collimated by an aspheric lens (Thorlabs C280TME-B) and reflected off two dielectric mirrors (Thorlabs BB1-E03-10) to provide four degrees of freedom for alignment. It then goes through 90:10 (reflection:transmission) non-polarizing beamsplitter (Thorlabs BS029 90:10), a mechanical shutter (Thorlabs SH05 with Thorlabs SC10 controller), a focusing lens (Thorlabs C220TME-B), and reaches the APD.

The continuous-wave annealing laser LD2 is coupled to a $200~\micro\meter$ diameter core MM fiber (RPMC Lasers OAL-200/220/245). The annealing laser beam is collimated by an aspheric lens (Thorlabs C220TME-B) and reflected at the 90:10 non-polarizing beamsplitter. The beams of the annealing and signal lasers overlap so that both can be focused on the same spot. A power meter (PM2; Thorlabs S142C) located in the transmission arm of the annealing laser allows us to accurately control the annealing power at the APD.

A camera (Canon 7D with macro lens MP-E $65~\milli\meter$ f/2.8 1--5x) and a light-emitting-diode photography illuminator are mounted beside the laser setup. The XYZ translation stage, on which the detector module is mounted, enables us to move the samples between the laser setup and the camera.

\subsection{Cold-temperature characterization apparatus}

A separate apparatus, shown in \cref{fig:experimental-setup}(b), is built to measure the dark count rate, relative changes in $P_\text{de}$, $\Delta t_\text{jitter}$, and afterpulsing probability at $-80~\celsius$. The low temperature significantly suppresses thermally excited dark counts \cite{anisimova2015}. The detector module is extracted from the laser annealing apparatus and placed inside a cooling chamber (Delta Design 9023) at $-80~\celsius$.

The apparatus necessary to measure the absolute $P_\text{de}$ cannot fit in the cooling chamber; consequently, we measure $P_\text{de}$ relative to a reference by sending diverging weak coherent pulses (WCPs) from the end of a fiber onto the APDs. Each APD's active area is sufficiently small that it receives approximately uniformly distributed light intensity, despite the overall Gaussian profile of the incident optical mode. We use a $780~\nano\meter$ laser LD3 (PicoQuant LDH 8-1) as the WCP source, pulsed at $40~\mega\hertz$ with full width at half maximum (FWHM) of $188~\pico\second$. The laser pulses are split by a 50:50 fiber beamsplitter (Thorlabs FC780-50B-FC). One output port is connected to a digital variable attenuator (OZ Optics DA-100-35-770/830/850-5/125-5-40-LL) and sends the laser pulses through a fiber to the APDs placed inside the cooling chamber. The other output port is connected to an optical-to-electrical converter (LeCroy OE455, DC--$3.5~\giga\hertz$) and the oscilloscope. Having oscilloscope access to both input laser pulses and output avalanche pulses allows us to measure $\Delta t_\text{jitter}$ (which requires using a pulsed laser, as opposed to continuous-wave such as LD1).

\section{Methods}
\label{sec:measurement}

\subsection{Laser annealing}

To perform laser annealing, the detector module is positioned to ensure that the high-power annealing laser beam is focused on the active area of an APD. Next, the desired annealing power is set by monitoring the power meter PM2 with the shutter closed. We then open the shutter and laser-anneal the APD for $60~\second$, immediately afterwards closing the shutter and letting the device cool down to the room temperature for another $60~\second$. We then perform characterization.

For most samples, we perform multiple stages of laser-annealing and characterization, with laser power increased between each stage. To determine whether this stepwise laser annealing process has any additional effects, we apply only a single-shot power of $1~\watt$ to SLiK-4 and SLiK-5 for comparison ($1~\watt$ is chosen based on observed results of the first three SLiKs to ensure a dark count rate reduction). Similarly, C30902SH-2 is laser-annealed at two specific powers, chosen based on C30902SH-1's observed results.

The temperatures reached by the APDs during laser annealing are of interest because, for some temperature ranges, alternative heating methods may be more practical to implement on a satellite (e.g., using an electric heater). We can measure the temperature of SLiK samples using their integrated thermistor (mounted on the cold plate of TEC, close to the APD). The temperature of SLiK-1's thermistor is recorded at the end of the $60~\second$ exposure, for most annealing powers. According to our measurements (see \cref{app:SLIK-thermal-resistance}), the thermal resistance between the photodiode chip and thermistor is negligible; thus, the temperature reading by the thermistor provides an accurate reading of the temperature reached by the APD during laser annealing.

\subsection{Characterization}
\label{subsec:charac}

All nine samples' parameters are measured at $-80~\celsius$ in the cold-temperature characterization apparatus, while SLiKs are also characterized at $-30~\celsius$ (reached using their built-in TECs) in the laser annealing apparatus.

\noindent \textbf{\textit{Breakdown voltage.}} We measure $V_\text{br}$ by observing avalanche pulses on an oscilloscope. As $V_\text{bias}$ is gradually increased from $0~\volt$, avalanche pulses begin to appear when $V_\text{br}$ is reached. We estimate the accuracy of this $V_\text{br}$ measurement to be better than $\pm0.3~\volt$. This measurement is performed at the start of every full characterization. Both C30902SH samples and SLiK-1, SLiK-2, and SLiK-3 are subsequently biased $+14~\volt$ above $V_\text{br}$, whereas SLiK-4, SLiK-5, and both SAP500S2 samples are biased $+20~\volt$ above $V_\text{br}$. The choice of $V_\text{excess}$ is arbitrary; however, we change $V_\text{excess}$ values to explore its effects on laser-annealed APDs.

\noindent \textbf{\textit{Dark count rate.}} We record dark counts with the time-tagging unit for $500~\second$. The mean dark count rate is calculated from the collected data. We define the dark count rate reduction factor as the ratio between the initial mean dark count rate and the mean dark count rate after laser annealing.

\noindent \textbf{\textit{Photon detection efficiency.}} To measure relative $P_\text{de}$ at $-80~\celsius$ in the cold-temperature characterization apparatus using diverging WCPs, we take the mean count rate over $500~\second$. Following the completion of laser-annealing and measurements of the two C30902SH samples, it was discovered that the output power of LD3 varies over time. As the C30902SH samples have undergone laser annealing, measurements unfortunately cannot be retaken. Consequently, for all other samples, we normalize the count rate by the average laser power (measured at the oscilloscope immediately before and after a measurement). We then calculate the relative change in $P_\text{de}$ by subtracting the mean dark count rate and then normalising against the initial photon count rate. 

For SLiKs we determine the absolute $P_\text{de}$ while operating at $-30~\celsius$ in the laser annealing apparatus. The continuous-wave signal laser LD1 is applied, with its photon rate calibrated via loss and laser power measurements in our optical setup. Then, we measure a mean count rate over $500~\second$ using the time-tagging unit. Subtracting the mean dark count rate from the mean count rate and dividing the difference by the calibrated input photon rate gives $P_\text{de}$.

Testing within the laser annealing apparatus also allows us to scan $P_\text{de}$ at various points over the active area of the SLiK devices. We measure the signal laser's illumination spot size using a beam profiler---the beam waist is ${\sim}20~\micro\meter$ (FWHM). We thus step the translation stages in $10~\micro\meter$ increments, such that $P_\text{de}$ of the active area is fully and tightly covered. We scan a square region that covers the active area, with appropriate pauses at each step for vibration dissipation.

\begin{figure}
\centering
\includegraphics[width=0.73\columnwidth]{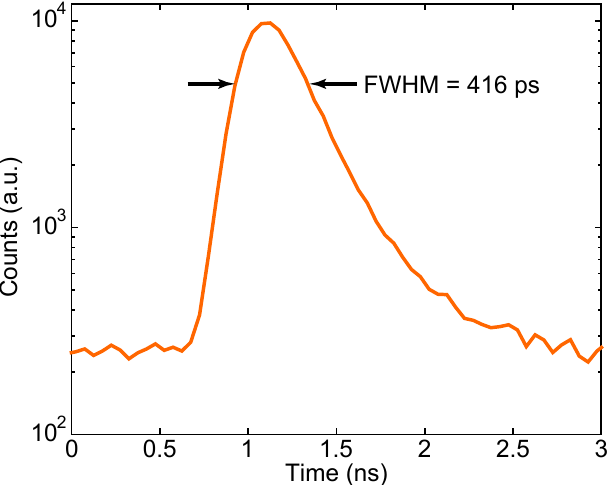}
\caption{\label{fig:jitter}{\bf Timing jitter of SAP500S2-2.} We measure the relative time difference between laser pulses and APD output pulses over at least $10^5$ output pulses, and plot these as a histogram. $\Delta t_\text{jitter}$ is the width of the pulse peak.}
\end{figure}

\noindent \textbf{\textit{Timing jitter.}} To measure $\Delta t_\text{jitter}$, both the laser output pulses and the APD's avalanche pulses are connected to the oscilloscope. We then plot a histogram of the relative time difference between these two signals over at least $10^{5}$ avalanche pulses (see example in \cref{fig:jitter}), and measure the FWHM.

\begin{figure}[h]
\centering
\includegraphics[width=0.7\columnwidth]{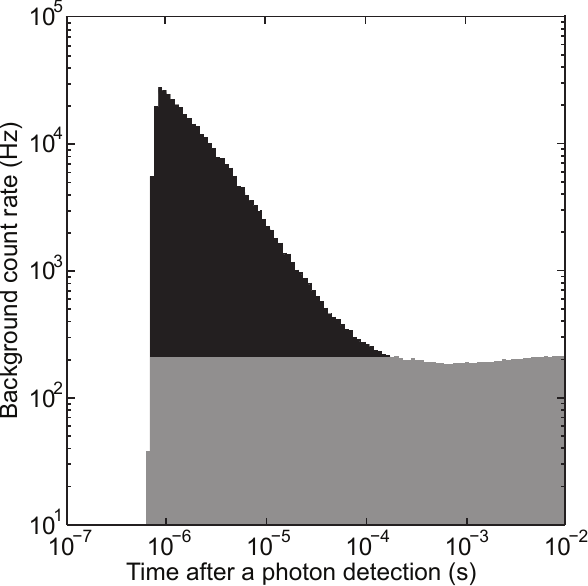}
\caption{\label{fig:afterpulsing}{\bf Afterpulsing histogram of SAP500S2-2.} The histogram of the time difference between APD output pulses shows the APD's dead time from 0 to roughly $0.8~\micro\second$, followed by a recharge time of ${\approx}150~\nano\second$. The following peak is formed of afterpulses. The count rates then settle down to the background count rate of ${\approx}200~\hertz$. A possible reason for a slight increase in the count rate at the end is the high voltage supply rebounding after being slightly sagged by the charge being drawn by the single avalanche. The black shaded area is the afterpulsing probability.}
\end{figure}

\noindent \textbf{\textit{Afterpulsing probability.}} Afterpulsing probability is calculated from timestamped counts. We typically use dark counts, but if the dark count rate for an APD is too low, we use a dim light pulsed at $100~\hertz$ to facilitate a measurement. The extra light increases the background count rate by a few $\hertz$ (otherwise, the background count rate tracks the APD's dark count rate), and doesn't influence the afterpulsing distribution.

For every detection event, our software adds all subsequent detection events occurring up to $10^{-2}~\second$ later to a histogram, with exponentially growing time bins \cite{anisimova2015-qcrypt}, as shown in \cref{fig:afterpulsing}. Unlike the standard autocorrelation method, this improved analysis prdocues a plot that converges to the background count rate, instead of following an exponential decay. Using exponentially increasing bin sizes filters out statistical fluctuations in the tail and also resolves the fast changing avalanche peak. 

The shape of this histogram (\cref{fig:afterpulsing}) presents four features: an APD's dead time, its recharge time, its trapped-electron time constants, and the background count rate. In \cref{fig:afterpulsing}, the dead time begins immediately after a photon detection (time 0) and ends when counts begin to reappear at roughly $0.8~\micro\second$. The recharge time is the time it takes for the count rate to reach the peak value after the dead time. Trapping time constants can be found by fitting the exponentially decaying slope of the peak. The plot levels off to the background count rate and the black shaded area is the afterpulsing probability. 

For a quantum communications protocol such as QKD, the afterpulsing peak contributes to the quantum bit error rate (QBER), and it is thus desirable to remove the peak by extending the dead time out to the flattened region \cite{hadfield2009, eisaman2011, cova1996}. In practice, this can be performed by discarding all counts before the end of a user-selected dead time in post-processing \cite{yoshizawa2002, yoshizawa2003} or by using an active quenching circuit \cite{sultana2016-qcrypt}. Such additional dead time, however, limits the maximum detection rate \cite{hadfield2009, cova2004}. Because long distance transmissions reduce the expected detection rate \cite{bourgoin2015}, the amount of additional dead time could be optimized to balance the QBER with the detection rate to maximize the final key rate \cite{anisimova2015-qcrypt, yoshizawa2002, yoshizawa2003}.

\section{Results and discussion}
\label{sec:results}

Right half of \cref{tab:APDs-conditions-and-results} summarizes the laser annealing results for the maximum dark count rate reduction. Detailed results for each APD model follows. 

\begin{figure}
\centering
\includegraphics[width=\columnwidth]{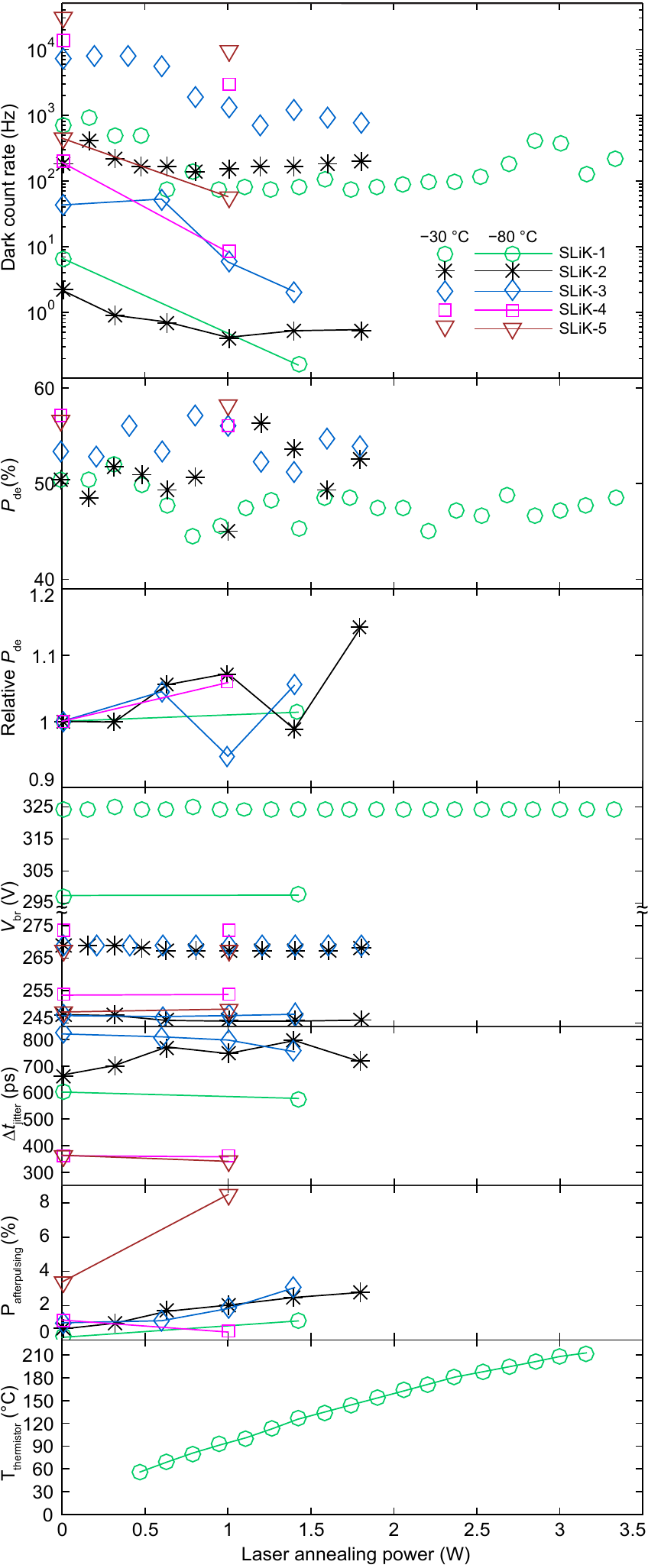}
\caption{\label{fig:SLIK-characterization}{\bf Characteristics of SLiK APDs after laser annealing.} Points at zero power show the initial characteristics before laser annealing. The samples have been characterized at $-30~\celsius$ after every annealing stage (data represented by points without lines), and additionally characterized at $-80~\celsius$ after some of the steps (points connected with lines).}
\end{figure}

\noindent \textbf{\textit{Excelitas SLiK.}} \Cref{fig:SLIK-characterization} shows characterization results for the five SLiKs. The maximum dark count rate reduction ranges from 1.3 to 10 times when operating at $-30~\celsius$, and from 5.3 to 41.7 times at $-80~\celsius$. Importantly, the SLiKs that were previously thermally annealed exhibit further dark count rate reduction after laser annealing. SLiK-4 and SLiK-5, after a single exposure at $1~\watt$, still show a significant decrease in the dark count rate. $1~\watt$ may not give the highest reduction factor but it shows that the single exposure and stepwise annealing processes induce comparable dark count rate reduction factors. Thus, the stepwise process can be used on a SLiK to find the optimal annealing power and a single exposure at this power on other SLiKs under similar conditions should achieve comparable dark count rate reduction.

\begin{figure}
\centering
\includegraphics[width=\columnwidth]{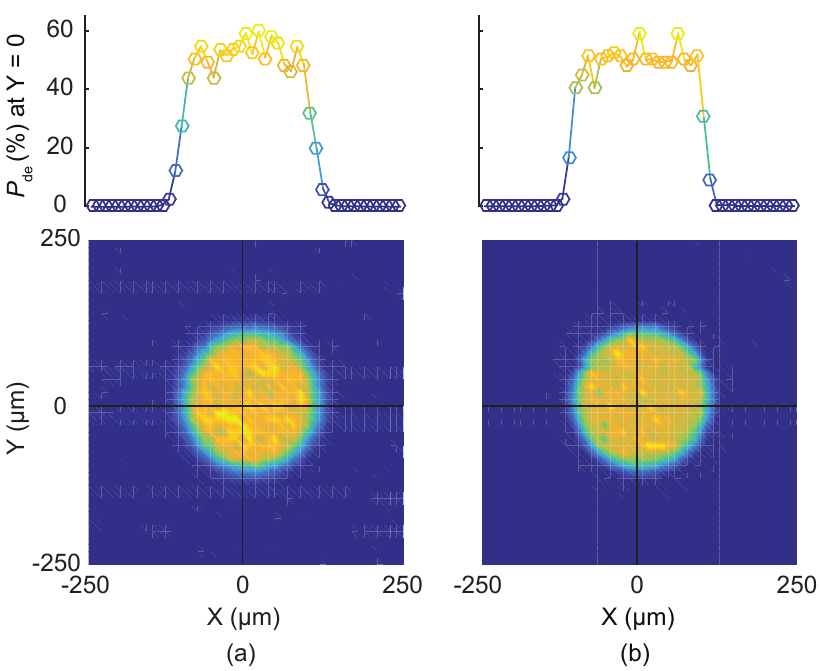}
\caption{\label{fig:SLIK-efficiency-MATLAB}{\bf Detection efficiency scan of SLiK-1.} (a) Prior to annealing, and (b) after $3.3~\watt$ annealing. The spatial profile is essentially unchanged, proving that focused high-power laser illumination does not degrade photosensitivity in the active region of the APD for the power range we tested. The plots at the top are the cross sections at Y = 0.}
\end{figure}

$P_\text{de}$ of each SLiK did not change significantly, and a spatial scan in \cref{fig:SLIK-efficiency-MATLAB} shows that photosensitivity across the active area is not altered by high-power laser annealing. $V_\text{br}$ and $\Delta t_\text{jitter}$ also do not fluctuate much after laser annealing. SLiK-4's and SLiK-5's $\Delta t_\text{jitter}$ are lower than those of other SLiKs due to the $6~\volt$ difference in $V_\text{excess}$ \cite{cova1996}.

The afterpulsing probability results are interesting. Proton radiation mainly causes displacement damage in APDs \cite{srour1988}. Highly energetic protons displace atoms from their lattice structures, producing extra energy levels in the bandgap (defects). The defects near the mid-bandgap contribute to thermally generated dark counts, while those near the conduction band are called trap levels that cause afterpulses. After proton irradiation, SLiKs' dark count rate significantly increased but the afterpulsing probabilities are still low \cite{anisimova2015}, implying that the displacement damage by $106~\mega\electronvolt$ protons at the depletion region mainly produces thermal generation centers \cite{srour1988}. Although the dark count rate is reduced in all samples after laser annealing, the afterpulsing probability increases in most of them. This also implies that laser annealing not only removes thermal generation centers but it can also create extra trap levels simultaneously. The increased afterpulsing probability can be handled by imposing additional dead time (see \cref{subsec:charac}).

When the high-power laser is applied to SLiK-1, its thermistor temperature rapidly increases in the first $30~\second$. The temperature continues to rise at a slower rate in the second half of the $60~\second$ exposure until it reaches the peak temperature in the last 2--3$~\second$. The peak thermistor temperature at each annealing power is plotted in \cref{fig:SLIK-characterization}. The SLiKs we have tested experience a significant dark count rate reduction at annealing power of $1~\watt$. Assuming the thermal resistance between SLiK-1's thermistor and the APD is negligible (see \cref{app:SLIK-thermal-resistance}), $1~\watt$ of power anneals the APD at peak temperature of ${\sim}90~\celsius$. The peak temperature is reached only in the last few seconds, but we speculate that it determines the dark count rate reduction factor. If this is the case, our results imply that laser-annealing irradiated SLiKs at the peak temperature of ${\sim}90~\celsius$ for $60~\second$ leads to a higher dark count rate reduction than thermally annealing them at $100~\celsius$ for two hours \cite{anisimova2015} (see \cref{tab:APDs-conditions-and-results}). The observed discrepancy in the dark count rate reduction suggests that light may play a greater role than simply providing the heat energy in improving irradiated SLiKs' performance.

\begin{figure}
\centering
\includegraphics[width=\columnwidth]{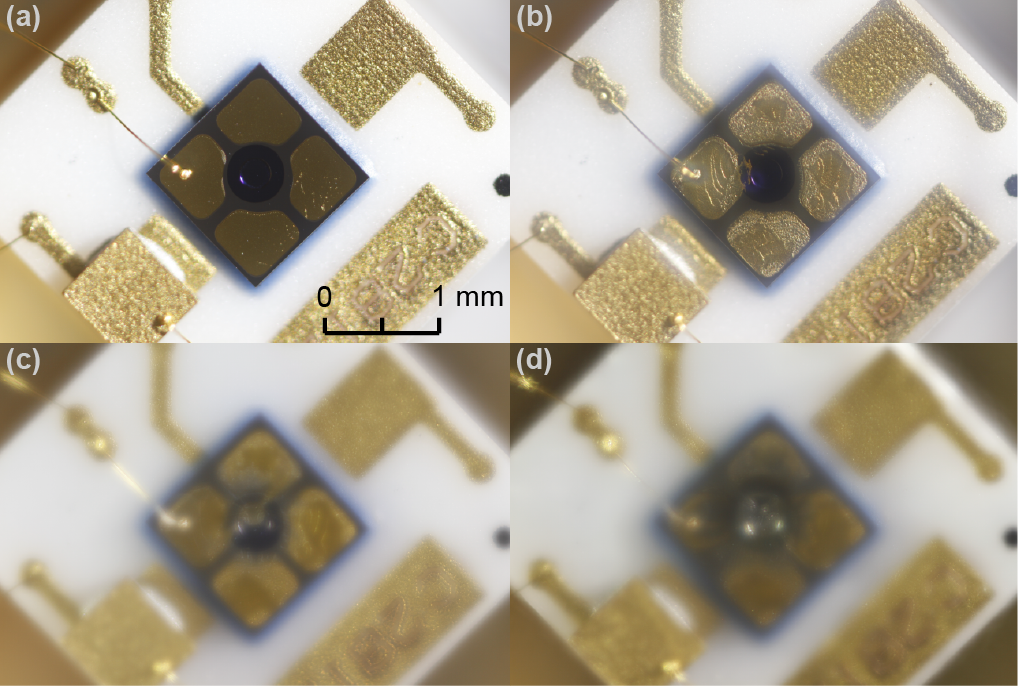}
\caption{\label{fig:SLIK-pictures}{\bf Progressive destruction of SLiK-1.} The sample is exposed to laser power of (a) $3.5~\watt$, (b) $5.2~\watt$, (c) $7.7~\watt$, and (d) $9.0~\watt$. The photodiode stopped working as a single-photon detector after annealing at $3.5~\watt$. At higher powers, the package window becomes foggy, and the gold plating surrounding the active area of the APD begins to melt and flow into the active area.}
\end{figure}

For SLiK-1, we continue to test at higher powers. \cref{fig:SLIK-pictures} shows the gradual change in the appearance of SLiK-1's active area. \cref{fig:SLIK-pictures}(a) does not show any visible physical destruction, but the device has already stopped working as a single-photon detector after annealing at $3.5~\watt$. In \cref{fig:SLIK-pictures}(b)--(d), damage becomes visible. It appears that an epoxy layer between two ceramic plates has boiled and condensed on the package window, causing it to become opaque. (SLiK's detailed mechanical structure can be seen in Fig.~5 of Ref.~\onlinecite{sauge2011}.)

\begin{figure}
\centering
\includegraphics[width=\columnwidth]{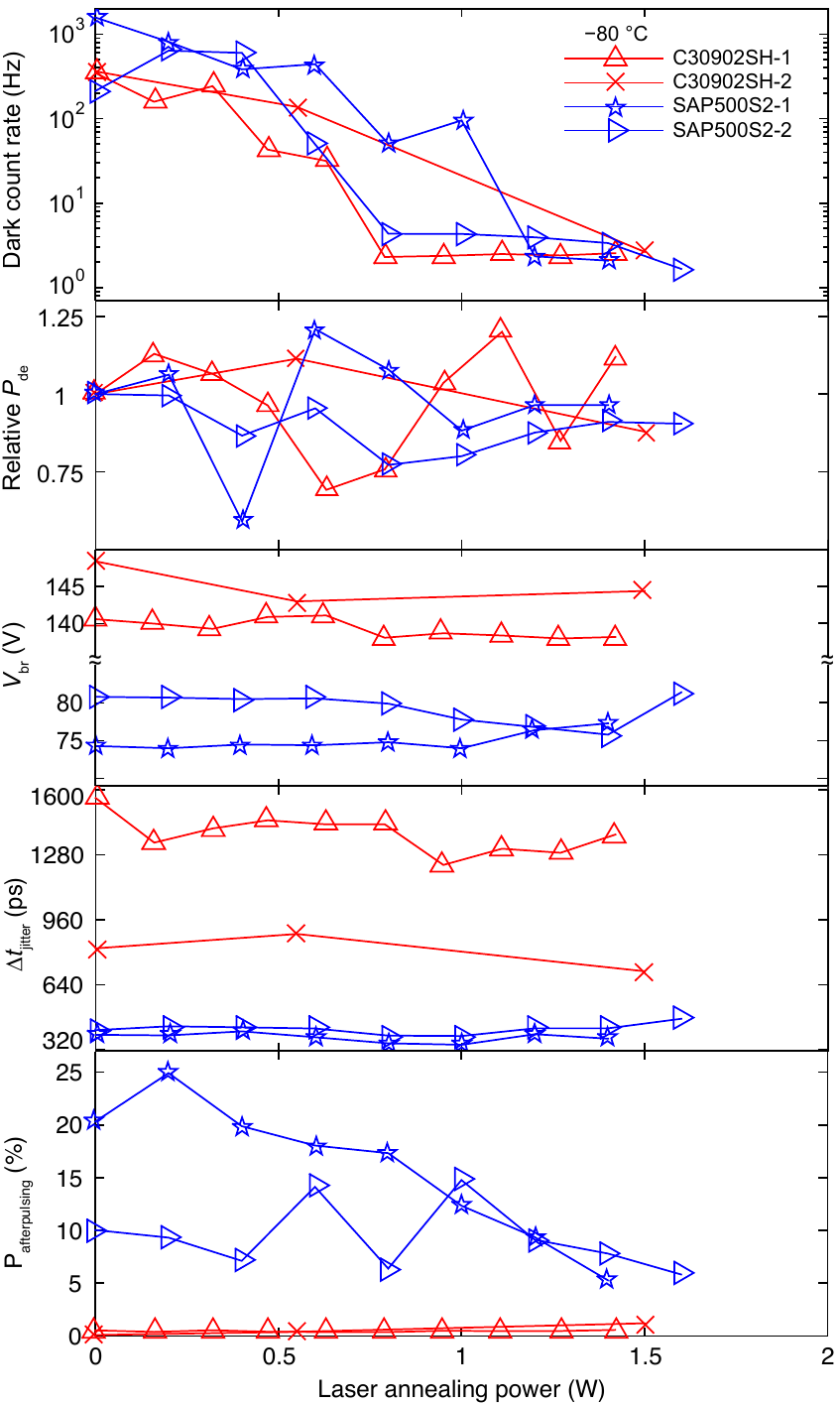}
\caption{\label{fig:SAP500/C30902SH-characterization}{\bf Characteristics of C30902SH and SAP500S2 APDs after laser annealing.} Points at zero power show the initial characteristics before laser annealing. Each point shows a measurement after successive laser illumination power.}
\end{figure}

\noindent\textbf{\textit{Excelitas C30902SH.}} \Cref{fig:SAP500/C30902SH-characterization} shows characterization results for the two irradiated C30902SHs. Similar to the SLiKs, C30902SH-1 is laser-annealed and characterized at multiple stages, but C30902SH-2 is treated at two specific powers, chosen based on C30902SH-1's results. The samples experience the maximum dark count rate reduction of 150 times and 137 times at $-80~\celsius$. Again, the stepwise process does not show any additional improvements on the dark count rate when compared to the single exposure process.

Relative $P_\text{de}$ measurement for C30902SHs is inaccurate due to the fluctuation of the laser output power (which we normalize out for measurements of the other devices; see \cref{subsec:charac}). Due to the variation observed, it is difficult to draw any conclusions. $V_\text{br}$ and $\Delta t_{jitter}$ do not change significantly. Afterpulsing probabilities are negligible in C30902SHs, which also implies that thermal generation centers are the main contributors to dark counts after irradiation. We stop testing these C30902SHs before they show signs of damage---testing non-irradiated C30902SHs at higher powers was done in Ref.~\onlinecite{bugge2014}.

\noindent\textbf{\textit{Laser Components SAP500S2.}} Unlike SLiKs, SAP500S2s cannot withstand high-power illumination. SAP500S2-1 and SAP500S2-2 fail to work as single-photon detectors after laser annealing at $1.6~\watt$ and $1.8~\watt$, respectively. Both SAP500S2s (see \cref{fig:SAP500/C30902SH-characterization}) exhibit the maximum dark count rate reduction just before they stop working as a single-photon detector (the reduction factor of 758 times for SAP500S2-1 and 128 times for SAP500S2-2). SAP500S2s have the highest dark count rate after proton irradiation \cite{anisimova2015}, but the lowest dark count rates after laser annealing (${\approx}2~\hertz$) are close to those of other APD models we have tested. 

Relative $P_\text{de}$ varies for different laser annealing powers. Unlike C30902SH measurements, the photon count rate here is normalised to the laser source's average power for every characterization, so the observed variation is real. To avoid $P_\text{de}$ fluctuation and achieve sufficiently low dark count rate, one should perform laser annealing around $1~\watt$.

$V_\text{br}$ seems to decrease slightly, and then increase just before each SAP500S2 stops working as a single-photon detector. An increase in $V_\text{br}$ may be an indication that the maximum laser annealing power is reached. $\Delta t_\text{jitter}$ remains almost constant and it is low compared to C30902SHs because SAP500S2s are biased an extra $6~\volt$ (for a $V_\text{excess} = 20~\volt$) above a significantly lower $V_\text{br}$. 

While SLiKs and C30902SHs display a marginal increase in afterpulsing probability as the laser power increases, the afterpulsing probability is reduced in SAP500S2s (afterpulsing probability did not change much after proton irradiation \cite{anisimova2015}). Although these reductions are insignificant compared to the highest dark count rate reduction factor in \cref{tab:APDs-conditions-and-results}, the results imply that laser annealing reduces both trap levels and thermal generation centers at the same time. The overall results suggest that the materials used to manufacture SAP500S2s are more susceptible to the high energy damages induced by protons and the annealing laser.

\section{Conclusion}
\label{sec:conclusion}

Our results demonstrate that laser annealing can remedy low Earth orbit radiation damage for three different types of APDs. The dark count rate of all samples, including the samples that were previously thermally annealed, is greatly reduced. This suggests that laser annealing is a more effective method. Fluctuations in other photon counting parameters should not degrade the performance of quantum communications applications. We speculate that laser annealing heals crystal lattice structure defects, thermal generation centers in particular, created by proton radiation. By employing this effect, the lifespan of a quantum satellite may be lengthened.

\section*{Author's contributions}
    J.G.L.\ performed the experiments, conducted data analysis, and wrote the manuscript with contributions from all authors. E.A.,\ B.L.H.,\ J.-P.B.,\ T.J.,\ and V.M.\ provided the samples and assisted with experimental setup and data analysis. T.J.\ and V.M.\ supervised the study.

\begin{acknowledgments}
We thank those authors of Ref.~\onlinecite{anisimova2015} not listed in the present paper, namely M.~Cranmer, E.~Choi, D.~Hudson, L.~P.~Piche, and A.~Scott, for assisting with the detector sample irradiation campaign. We thank Excelitas for discussions and for providing selected APD samples. We thank T.~Graham and P.~Kwiat for an independent experimental confirmation of the laser annealing effect on an irradiated APD. This work was supported by the Canadian Space Agency, Industry Canada, CFI, NSERC (programs Discovery and CryptoWorks21), Ontario MRI, and the U.S.\ Office of Naval Research.
\end{acknowledgments}

\appendix
\renewcommand\thesection{} 
\section{Measurement of laser annealing temperature}
\label{app:SLIK-thermal-resistance}

\begin{figure}[h]
\centering
\includegraphics[width=0.87\columnwidth]{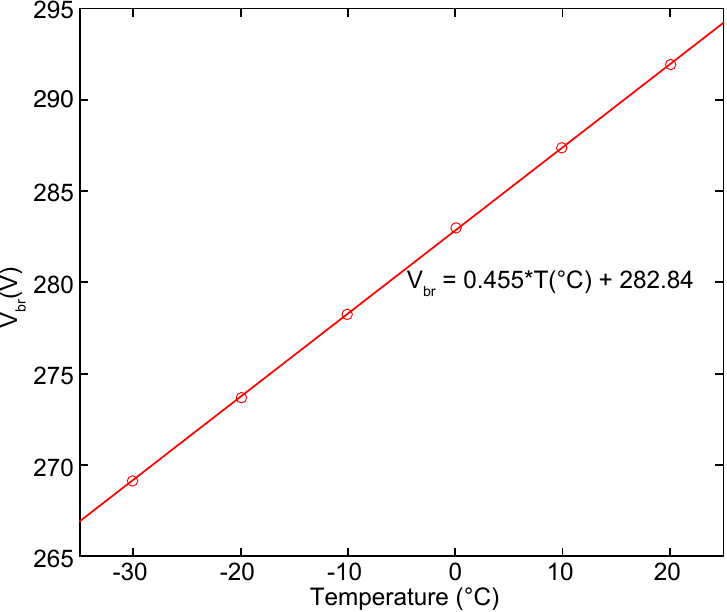}
\caption{\label{fig:APD-breakdown-voltage-and-temperature}{\bf Breakdown voltage ($\bm{V_\text{br}}$) of SLiK-2 at various temperatures.} The relationship between $V_\text{br}$ and temperature is linear. The coefficient of determination, $R^{2}$, is $0.99994$.}
\end{figure}

We measure the thermal resistance between the APD and the thermistor of SLiK-2, in order to estimate temperatures achieved by laser annealing. The thermistor mounted on the cold plate of the TEC (see Fig.~5 in Ref.~\onlinecite{sauge2011}) allows us to measure the cold plate's temperature during the laser annealing process. Thus, by finding the thermal resistance between the APD and the cold plate, we can predict the actual temperature produced by each annealing power.

Alternatively, measuring $V_\text{br}$ provides an indirect measure of an APD's temperature. This method is convenient because we can easily measure $V_\text{br}$ with the experimental setup. Before measurements, we find the relationship between $V_\text{br}$ and temperature by setting the TEC controller to temperatures from $-30~\celsius$ to ${+}20~\celsius$ in $10~\celsius$ increments and measuring $V_\text{br}$ at each point. The fit line in \cref{fig:APD-breakdown-voltage-and-temperature} shows that $V_\text{br}$ linearly depends on temperature. 

\begin{table}
	\centering
	\setlength{\textfloatsep}{0.1cm}
	\setlength{\floatsep}{0.1cm}
	\centering
	\caption{\label{tab:APD-thermal-resistance}{\bf Measurement of thermal resistance between APD and cold plate.}}
		\renewcommand{\arraystretch}{1.45}

   \begin{tabular*}{\linewidth}
   {>{\centering\arraybackslash}m{0.18\linewidth}
    	 >{\centering\arraybackslash}m{0.17\linewidth}
    	 >{\centering\arraybackslash}m{0.15\linewidth}
    	 >{\centering\arraybackslash}m{0.22\linewidth}
    	 >{\centering\arraybackslash}m{0.2\linewidth}}
    \hline\hline
	\textbf{$\bm{V_\text{bias}}$ ($\bm{\volt}$)} & \textbf{$\bm{V_\text{br}}$ ($\bm{\volt}$)} & \textbf{$\Delta$T ($\bm{\celsius}$)} & \textbf{Power dissipation ($\bm{\milli\watt}$)} & \textbf{Thermal resistance ($\bm{\kelvin\per\watt}$)} \\

	\hline

    $340$ & $294.65$ & $-0.16$ & $33.4$ & $-4.8$\\
    $350$ & $295.05$ & $-0.44$& $40.5$ & $-10.8$\\
    $360$ & $295.82$ & $0.11$& $47.4$ & $2.3$\\
    $370$ & $296.74$ & $0.98$& $54.3$ & $18.0$\\
    $380$ & $297.10$ & $0.54$& $61.5$ & $8.8$\\

	\hline\hline
  \end{tabular*}
\end{table}

For the thermal resistance measurement, we turn off the TEC current but keep measuring thermistor temperature. We also vary the bias voltage over the range of $50~\volt$ to $90~\volt$ above $V_\text{br}$ at room temperature. Such high bias voltages cause a high dark count rate, resulting in a constant current flow in the circuit. Consequently, the voltage across the APD is approximately $V_\text{br}$ because of continuous quenching process, and $V_\text{br}$ increases at high bias voltages owing to higher heat dissipation. When applying each bias voltage, we wait for the thermistor temperature to stabilize, and measure the voltage across the $1~\kilo\ohm$ readout resistor in the passive quenching circuit (replacing R2 in \cref{fig:detector-module} for more accurate avalanche current measurement). From this voltage value, we deduce $V_\text{br}$ and the APD's power dissipation. Power dissipation of the APD is calculated by multiplying the avalanche current (on the order of a few hundred ${\micro\ampere}$) and $V_\text{br}$. Using the linear relationship found in \cref{fig:APD-breakdown-voltage-and-temperature}, we find the APD's temperature and compare it to the thermistor temperature. The temperature difference divided by the APD's power dissipation gives us the thermal resistance. \cref{tab:APD-thermal-resistance} summarizes thermal resistance test results. Although the results are somewhat noisy, making it difficult to find the exact thermal resistance, it is clear that the thermal resistance is sufficiently small that the thermistor temperature approximately matches that of the APD itself. This assumption is used to estimate laser annealing temperature in conjunction with SLiK-1's thermistor temperature measurements during laser annealing procedure. 

\bibliography{bibtex_library} 

\end{document}